\documentclass[%
 reprint,
 showkeys,
 amsmath,amssymb,
 aps,
pra,
]{revtex4-2}

\usepackage{graphicx}
\usepackage{dcolumn}
\usepackage{bm}
\usepackage{siunitx}

\usepackage{multirow}
\usepackage{xfrac}

\begin{document}

\preprint{APS/123-QED}

\title{Towards noble gas quantum optical magnetometry using direct ultraviolet detection}

\author{James Maldaner}
 \email{maldaner@ualberta.ca}
 \affiliation{Department of Physics, University of Alberta}
\author{Gil Porat}%
 \email{gporat@ualberta.ca}
\affiliation{%
 Department of Physics, University of Alberta\\
 Department of Electrical and Computer Engineering, University of Alberta
}%

\date{\today}

\begin{abstract}
We analyze a novel approach for quantum magnetic sensing via optical detection of nuclear spin precession in a noble gas. The detection is carried out with an ultraviolet frequency comb laser, so no alkali vapor is required to be mixed with the noble gas during the sensing stage, allowing for hours-long spin relaxation time at room temperature. Our analysis reveals that combining high laser power and low laser intensity noise is crucial to achieving and exceeding state-of-the-art magnetometric performance, including that of cryogenic magnetometers. We find operational regimes where either fundamental or technical noise dominate, which directs experimental effort towards increasing laser power or lowering laser intensity noise, respectively. We also show that optimizing beam size is key to achieving optimal sensitivity. Our calculations predict that a record-breaking sensitivity of $\SI{24}{aT/\sqrt{Hz}}$ is achievable in 10 minutes with a 10 W ultraviolet laser with a relative intensity noise of -100 dBc/Hz.
\end{abstract}

\maketitle

\section{Introduction}
Highly sensitive and precise magnetic sensing is crucial to a variety of applications \cite{Budker2007, Budker2013}. Examples include chemical analysis via nuclear magnetic resonance, biomedical imaging and sensing, fundamental physics tests, environmental sensing, and resource exploration. Some of the most notable applications that would benefit from sub-femtotesla/Hz$^{1/2}$ sensing include biomedical sensing~\cite{Murzin2020}, low-field nuclear magnetic resonance spectroscopy ~\cite{Jiang2021}, navigation~\cite{Zhai2022}, and fundamental physics research, such as our recently proposed search for quantum gravity effects \cite{Maldaner2024} and in the search for the neutron electric dipole moment~\cite{Pendlebury2015}.  These applications drive ever-increasing improvements in magnetometer sensitivity, which motivate using quantum systems to leverage coherent quantum phenomena.

The leading technological platforms for highly sensitive quantum magnetometers are superconducting quantum interference devices (SQUID), diamond nitrogen-vacancy (NV) centers, and atomic magnetometers. SQUIDs are capable of achieving a sensitivity of $\SI{150}{aT/\sqrt{Hz}}$ \cite{Storm2017}; however, they require cryogenic cooling since they are based on superconductors. This introduces technical challenges that limit their applicability. Diamond NV centers boast excellent properties such as room-temperature operation, optical readout, and all-solid-state composition. However, their sensitivity is limited to more than $\SI{1000}{fT/\sqrt{Hz}}$ \cite{Barry2020}, mainly due to their millisecond-scale spin lifetime. Finally, atomic magnetometers use a laser to detect the atomic spin precession in a magnetic field of an atomic agent that is spin-polarized by optical pumping. They have reached a sensitivity of $\SI{160}{aT/\sqrt{Hz}}$ \cite{Dang2010}, which is on par with SQUIDs and achieved in a heated mix of buffer gases and alkali vapor. Therefore, atomic magnetometers promise further sensitivity improvement with non-cryogenic detection.

The most sensitive atomic magnetometers utilize alkali atoms in the spin-exchange relaxation-free (SERF) regime \cite{Budker2007, Budker2013}. This regime is obtained when the rate of the alkali atoms' spin-exchange interaction is much greater than the precession (Larmor) frequency. The  rate of alkali-alkali spin exchange depends strongly on the physical conditions of the alkali vapor, mainly its density and temperature, where magnetic field strength and homogeneity determine the rate of spin-exchange relaxation. It has been established that spin-exchange relaxation is the dominant spin relaxation mechanism in alkali atomic magnetometers, significantly limiting their sensitivity. In the SERF regime, characterized by high alkali density at high temperatures and extremely low magnetic fields, spin-exchange relaxation is strongly suppressed. Under these conditions, spin relaxation time significantly exceeds $\SI{10}{ms}$ and has been shown to reach the minute scale  in a cell with specialized anti-relaxation coating \cite{Balabas2010}. However, suppression of spin-exchange relaxation requires careful zeroing of the magnetic field and is often sacrificed in favor of other considerations (such as avoiding $1/f$ noise). In either case, spin relaxation remains a significant limiting factor in the sensitivity of alkali atomic magnetometers.

Recently, proposals for a new paradigm of atomic magnetometers have emerged, which are based on optical access to the ground state of spin-polarized noble gas atoms \cite{Degenkolb2016, Mihara2016, Altiere2018}.  These proposals leverage the established technique of spin-polarization of noble gases via spin-exchange optical pumping (SEOP), which, together with gas purification, produces large amounts of spin-polarized and pure noble gas \cite{Walker1997, Nikolaou2013}. The main appeal of noble gas atoms is their extremely long spin relaxation time, which has been experimentally demonstrated to reach hours (e.g., for $^{129}\text{Xe}$ \cite{Kilian2007}). This property is facilitated by these atoms' complete electron shell structure, yielding a net-zero electronic spin in the ground state, leaving the nuclear spin as the only source of angular momentum. However, efficiently driving single-photon transitions from the ground state of noble gas atoms requires laser light in the vacuum to the extreme ultraviolet (UV) ($\SIrange{58}{147}{nm}$), which is extremely challenging to produce. To address this issue, two-photon excitation of $^{129}\text{Xe}$ at $\SI{252.5}{nm}$ or $\SI{256}{nm}$ has been considered \cite{Degenkolb2016, Mihara2016, Altiere2018}. Once the atoms have been excited using the two-photon transition, they decay by emitting an infrared (IR) photon, which is easily detected. While generating laser light in the deep UV is more straightforward than in extreme UV, it is still far from trivial. One proposed approach relies on the third-harmonic generation of a Ti:sapphire mode-locked laser \cite{Degenkolb2016}, and the other on the fourth-harmonic generation of a continuous-wave (CW) semiconductor laser \cite{Mihara2016, Altiere2018}. In both cases, useful power was limited to about $\SI{200}{mW}$, which was sufficient for spectroscopic measurement of a two-photon transition with tightly focused beams ($1/e^2$ beam radius $\ll \SI{1}{mm}$) in $\sim \SI{1}{mbar}$ of $^{129}\text{Xe}$ pressure, but so far there has been no demonstration of magnetometry, to the best of our knowledge. Furthermore, a limiting factor that has been overlooked in the above proposals is the effect of laser light intensity noise, commonly quantified as relative intensity noise (RIN)~\cite{shi2023}. This is the fractional variance in the laser intensity. RIN includes both technical and fundamental sources of noise, such as shot noise.

Here we analyze the sensitivity limits of a $^{129}\text{Xe}$ optical atomic magnetometer that does not use alkali vapour for detection. We show that a higher laser power than utilized to date is required for achieving and exceeding state-of-the-art magnetometric sensitivity and that it must be delivered with low RIN. We also show that optimizing the beam size is important to strike an optimal balance between spin projection noise and photon shot noise. Depending on operational parameters, we find optimal beam size ranging from 0.1 to 10 mm. To the best of our knowledge, this is the first time that the impact of laser intensity noise on alkali-free $^{129}\text{Xe}$ magnetometry is considered. We propose using a balanced detector (BD) scheme to reduce the impact of RIN by suppressing noise that is common to the excitation laser and the emitted fluorescence. We find that a sensitivity of $\SI{24}{aT/\sqrt{Hz}}$ (i.e., 6 times better than the current record) is at hand with a laser power of $\SI{10}{W}$, a RIN of $\SI{-100}{dBc/Hz}$, and a measurement time of $\SI{10}{min}$. Towards this, we propose using the fourth-harmonic of a high-power ytterbium-doped fiber frequency comb laser as the necessary UV laser light and discuss its feasibility.

\section{Principle of Operation and Basic Model}

\subsection{Operation and Experimental Layout}
\label{sec:operation}

\begin{figure}[htbp]
\centerline{\includegraphics[width=0.5\textwidth]{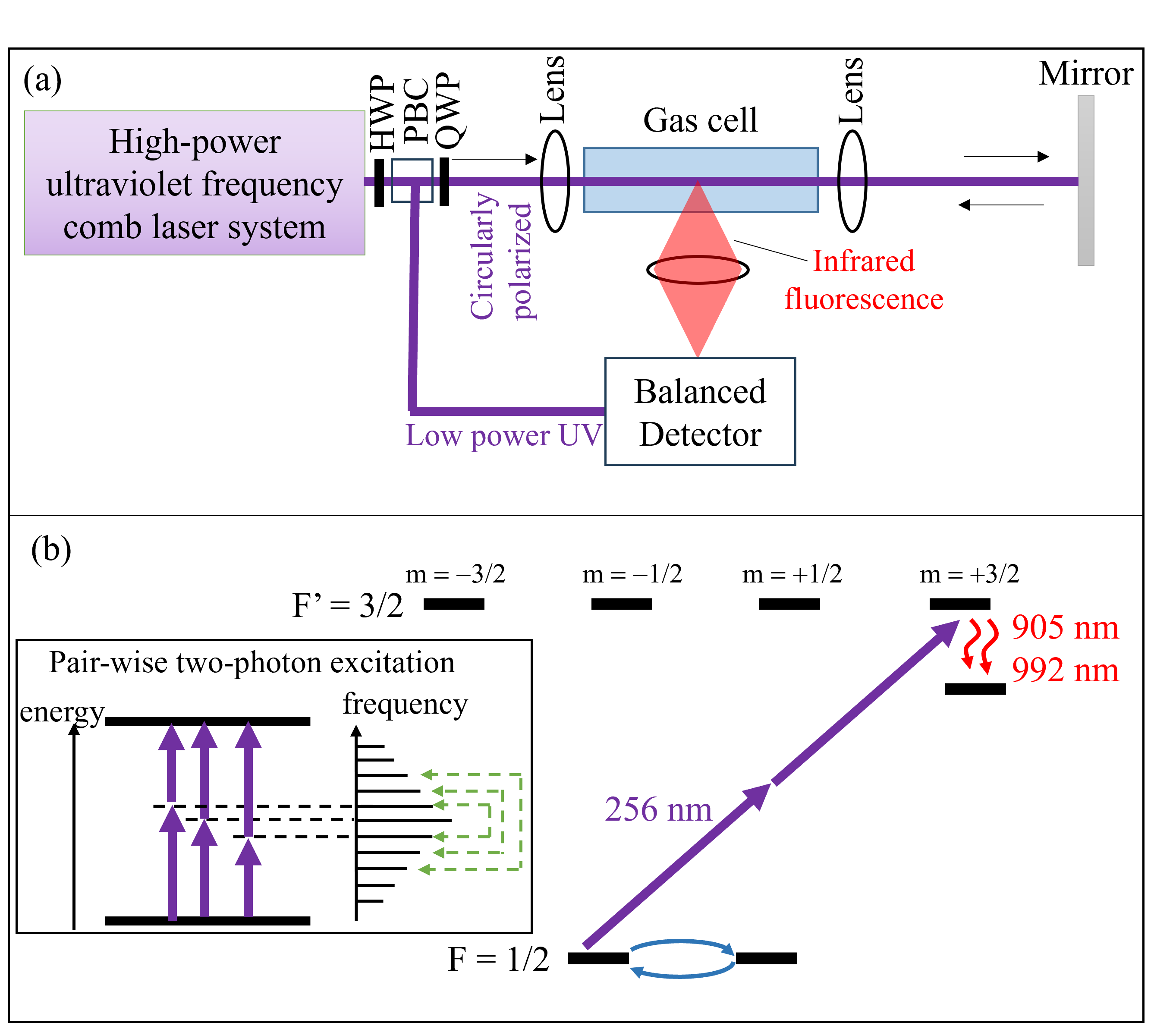}}
\caption{(a) Proposed experiment schematic for the DF magnetometry scenario. (b) Partial energy level diagram of $^{129}\text{Xe}$ and its proposed interaction with the laser. The inset illustrates the concept of pairwise two-photon excitation with a frequency comb.}
\label{fig:experiment}
\end{figure}

The scenario discussed here is based on Doppler-free (DF) two-photon frequency comb excitation \cite{Zhang2015, Picque2019} of $^{129}\text{Xe}$, as shown in Fig. \ref{fig:experiment} (a).
A high-power UV frequency comb laser beam is split using a half-wave plate (HWP) and a polarizing beam splitter (PBS). A small amount of the power ($ <\SI{10}{mW}$) is sent to a BD, and the remaining power is circularly polarized using a quarter wave plate (QWP) and sent to a xenon gas cell to facilitate two-photon excitation. This high-power portion is focused into the xenon gas cell, collimated, retroreflected, and focused again onto the same spot in the gas cell. Thus, counter-propagating beams illuminate the xenon atoms, and the Doppler shift of any atomic transition resonance frequency is eliminated (to first order). The fluoresced IR light is sent into the other half of the BD.  

The interaction between the laser light and $^{129}\text{Xe}$ atoms is illustrated in Fig. \ref{fig:experiment} (b). In xenon, as in all noble gases, the complete electron shell configuration means that there is net-zero electronic angular momentum in the ground state. Therefore, for $^{129}\text{Xe}$, the only angular momentum of the ground state comes from its nuclear spin of $I=1/2$, so it has only two magnetic sublevels. We use a quantization axis that is parallel to the direction of propagation of the laser beam. In our proposed scheme, the laser is tuned to the two-photon transition between the ground and $5 \text{p}^{5}\left( ^{2}\text{P}_{3/2} \right) 6 \text{p}^{2} \left[ 5/2 \right]_{2}, F^\prime=3/2$ state, corresponding to a laser wavelength of $\lambda_{UV} = \SI{256}{nm}$~\cite{vomHovel2023, Raymond1984, kroll1990, Saloman2004}. Furthermore, the laser is circularly polarized so that only the $m=-1/2$ magnetic sublevel of the ground state is dipole-allowed to be excited \cite{Bonin1984}. This transition satisfies the two-photon selection rules with $\Delta F=1$,  where the ground state has $F=1/2$, and $\Delta m_F=2$, see Appendix~\ref{app:two_photon_rules} for details. The excited state lifetime of $\SI{38}{ns}$ \cite{Whitehead1995} is much faster than the two-photon transition rate. The excited atoms decay to the $5 \text{p}^{5} \left( ^{2} \text{P}_{3/2} \right) 6 \text{s}^{2}[3/2]$ state by emitting infrared light \cite{Whitehead1995,Saloman2004}. Suppose a magnetic field is present in the volume occupied by the xenon gas, perpendicular to the quantization axis. In that case, the ground state population will undergo coherent oscillations between the two magnetic sublevels at the Larmor frequency \cite{Budker2007} (represented by the blue arrows in Fig. \ref{fig:experiment}  (b)) $\omega_B = \gamma_\text{Xe} B$, where $\gamma_\text{Xe}= 2\pi \times \SI{11.78}{MHz/T}$ \cite{Fan2016} is the gyromagnetic ratio of $^{129}\text{Xe}$, and $B$ is the magnetic field strength. Therefore, the excitation rate will oscillate accordingly, and so will the power of the detected infrared fluorescence, from which the magnetic field strength can be directly determined.  In principle, this scheme can be implemented using a continuous wave (CW) laser. Our choice of a frequency comb laser is motivated by practical considerations detailed in Appendix~\ref{app:laser}.

To be clear, driving the two-photon transition is not taken to facilitate optical pumping, i.e., it does not significantly alter the distribution of atomic population between different states. The initial out-of-equilibrium spin polarization is taken to be created a priori, e.g., via spin-exchange optical pumping, where subsequently the noble gas has been separated from the alkali atoms via cryogenic trapping (as commonly done in such systems~\cite{Khan2021}).  Here, the two-photon transition is used strictly for reading out the spin precession in the $^{129}\text{Xe}$ ground state.

Note that the experiment proposed here differs from our previous proposal for magnetometric measurement of quantum gravity effects \cite{Maldaner2024} in three ways. First, the scheme proposed here measures the absolute value of the magnetic field strength and not the difference in precession frequency between atoms traveling at different velocities. Second, it is in a DF geometry, whereas our previous work utilized Doppler broadening. Finally, in this paper, we consider the effects of RIN and detector noise and include a BD to mitigate the impact of RIN.

\subsection{Basic Model} \label{sec:basic_model}

Here we detail a basic model for the detected signal, i.e., the relation between the output current of the BD and the Larmor frequency. For now, we ignore noise, which we introduce in Section \ref{sec:sensitivity_model}.

Under the influence of a magnetic field, the atomic population oscillates between the ground state's two magnetic sublevels at frequency $\omega_B$. As the spins precess, the difference between the populations of these two magnetic sublevels oscillates and exponentially decays towards zero due to spin relaxation. In principle, both longitudinal and transverse spin relaxation contribute to this decay. However, transverse relaxation (due to dephasing) at rate $1/T_2$ dominates, since any effect contributing to longitudinal relaxation (at rate $1/T_1$) also contributes to dephasing.

The infrared fluorescence follows the population in one of the ground state sublevels, thus it is described by a decaying sinusoid with a nonzero mean value. The corresponding rate of photon emission is
\begin{equation}
    n_f(t)=\bar{n}_f\left[1 + P_{\text{pol}}\cos(\omega_B t)\exp\left(-{t}/{T_2}\right)\right],
    \label{eq:n_f(t)}
\end{equation}
where $0\leq P_{\text{pol}} \leq 1$ is the initial degree of spin polarization of the xenon gas, and $\bar{n}_f$ is the mean rate of infrared photon emission,
\begin{equation}
    \bar{n}_f = \frac{1}{2}W^{(2)}(\omega_0) N_{\text{Xe}}.
    \label{eq:n_f}
\end{equation}
Here, $W^{(2)}(\omega)$ is the two-photon transition rate per atom for incident light with an angular frequency $\omega$, $\omega_0$ is the two-photon resonance frequency, and $N_{\text{Xe}}$ is the number of atoms in the laser-gas interaction volume. The factor of $1/2$ accounts for the fact that when the atoms are unpolarized, each ground state's magnetic sublevel contains half the atomic population. 
The two-photon transition rate per atom for DF spectroscopy is~\cite{Raymond1984} 
\begin{equation}
    W^{(2)}(\omega_0)=24I_e^2\frac{\alpha_{\text{Xe}}}{\gamma},
    \label{eq:W2}
\end{equation}
where $I_e$ is the excitation intensity, $\alpha_{\text{Xe}}$ is the two-photon absorption coefficient and $\gamma$ is the two-photon transition linewidth.

We assume that the UV laser has a Gaussian beam where the power and average intensity are related through $I_e=P_{e}/\left( \pi w_0^2 \right)$, where $P_e$ is the average laser power incident on the xenon ensemble and $w_0$ is the beam radius at $1/e^2$ of its maximum intensity. We assume that the beam's confocal length greatly exceeds the interaction length, i.e., $2Z_R \gg L$, where $Z_R=\pi w_0^2 / \lambda_{UV}$ is the beam's Rayleigh length and $L$ is the interaction length. The interaction volume is thus $L \pi w_0^2$. The number density of $^{129}\text{Xe}$ atoms is given by the ideal gas law as $n_\text{Xe} = \eta_{^{129}\text{Xe}} p_\text{Xe} / \left( k_B T_\text{Xe} \right)$, where $\eta_{^{129}\text{Xe}}$ is the isotopic fraction of $^{129}\text{Xe}$ in the gas, $p_\text{Xe}$ is the gas pressure, $k_B$ is the Boltzmann constant, and $T_\text{Xe}$ is the xenon gas temperature. Therefore, the number of xenon atoms involved in the interaction is $N_{\text{Xe}} = \left( L \pi w_0^2 \right) n_{\text{Xe}}$.

The IR optical power incident on the detector is
\begin{equation}
P_{IR} = \hbar \omega _{IR}\epsilon_gn_f(t) ,
\label{eq:optical-power}
\end{equation}
where $\epsilon_g$ is the fraction of fluoresced IR photons collected into the photodetector and $\omega_{IR}$ is the average optical frequency of the fluoresced photons. The current of the IR photodiode is $i_{IR} = P_{IR}R(\omega_{IR})$, where
\begin{equation}
R(\omega) = \frac{e\epsilon_q(\omega)}{\hbar \omega}
\label{eq:responsivity}
\end{equation}
is the responsivity of each of the detector's identical photodiodes at optical frequency $\omega$. Here, $\epsilon_q(\omega)$ is the photodiode quantum efficiency at the same frequency and $e$ is the electron charge.
The total detection efficiency of each emitted IR photon is $\epsilon_d=\epsilon_g \epsilon_q(\omega_{IR})$. Using equations~\ref{eq:optical-power} and \ref{eq:responsivity} we find
\begin{equation}
    i_{IR}  =  e\epsilon_dn_f(t). 
    \label{eq:i_IR}
\end{equation}
We rewrite this as 
\begin{equation} \label{eq:IR_photocurrent}
    i_{IR} = \bar{i}_{IR} + \tilde{i}_{IR}(t)\exp(-t/T_2),
\end{equation}
where $\bar{i}_{IR}= e \epsilon_d\bar{n}_f$ is the DC component of the IR photodiode current and $\tilde{i}_{IR}(t)=e \epsilon_d\bar{n}_f P_{\text{pol}}\cos(\omega_B t)$ is the oscillating component of the IR photodiode current.

Next, we consider the low-power UV reference signal power, $P_r$, incident onto the detector. The corresponding photodiode current is
\begin{equation}
    i_r = P_{r}R(\omega_{UV})
    \label{eq:UV_photocurrent}
\end{equation}
and the BD output current is
\begin{equation}
    i_{BD} = i_r - i_{IR}.
\end{equation}
For optimal suppression of common mode noise, the BD should output zero current when the xenon ensemble is completely depolarized. The low-power UV reference can be set to fulfill this requirement, so that when $P_\text{pol} = 0$, we have $i_{BD} = 0 = (i_r - \bar{i}_{IR})$, therefore $i_r = \bar{i}_{IR} = e\epsilon_d\bar{n}_f$. We can then use Eq. \ref{eq:UV_photocurrent} to find the power required for the low power UV reference as $P_{r} = e\epsilon_d \bar{n}_f / R(\omega_{UV})$.

Under these conditions, the output current of the BD is
\begin{equation} \label{eq:BD_output}
    i_{BD}   = -e\epsilon_d\bar{n}_fP_\text{pol}\cos(\omega_B t)\exp(-t/T_2).
\end{equation}
As designed, there is no DC component in $i_{BD}$, and we have obtained an expression relating the BD's output current to the Larmor frequency.

\section{Analysis of Magnetometric Sensitivity} \label{sec:sensitivity_model}
In this section we first define magnetometric sensitivity and its relation to the error in the estimation of the Larmor frequency. Next, we provide a general expression for the impact of amplitude noise on frequency estimation. We use this expression extensively in the remainder of this section, where we analytically derive the Larmor frequency noise resulting from each mechanism we consider. We divide these mechanisms into two groups: fundamental noise (Section~\ref{sec:fund-noise}), which sets the ultimate limit on sensitivity, and technical noise (Section~\ref{sec:tech-noise}), for which we use realistic values in Sections~\ref{sec:parameters_opt} and~\ref{sec:results}.

\subsection{Definition of Magnetometric Sensitivity}
It is conventional to express magnetometric sensitivity in a manner that takes into account the measurement bandwidth (inverse of a single measurement time), i.e., the noise amplitude per unit bandwidth (in units of, e.g., $\SI{}{T/\sqrt{Hz}}$):
\begin{equation}
    \mathcal{S}_B = \sqrt{\frac{ T_m }{N_m}}\sigma_B = \sqrt{\frac{T_m}{ N_m}}\frac{\sigma_{\omega_B}}{\gamma_\text{Xe}},
    \label{eq:sensitivity}
\end{equation}
where $\sigma^2_B = \sigma^2_{\omega_B}/\gamma^2_\text{Xe}$ is the magnetic field variance, $\sigma_{\omega_B}^2$ is the variance in the precession frequency, $T_m$ is the duration of a single measurement, and $N_m$ is the number of times the measurement is repeated (for averaging).
To combine multiple noise sources, denoted $\sigma_{\omega_B,i}$, we add them in quadrature, i.e.,
\begin{equation}
    \sigma_{\omega_B}^2 = \sum_i \sigma_{\omega_B, i}^2.
    \label{eq:quad-add}
\end{equation}
\subsection{Decaying sinusoid frequency estimation error}

The BD output current is a decaying sinusoidal as given by Eq. \ref{eq:BD_output}. The variance in estimating its oscillation frequency $\omega_B$ is \cite{Swallows2013}
\begin{equation}
    \sigma_{\omega_B}^2 = \frac{6 \mathcal{N}_0^2 \left( e^{2T_m/T_2} - 1 \right) T_2}{\left|\text{max}(i_{BD})\right|^2 T_m^4}. 
    \label{eq:sigma_omega_B}
\end{equation}
For most noise sources considered below, the primary objective of each section is to determine the corresponding root-mean-square (RMS) noise spectral density, $\mathcal{N}_0$. Once this is obtained, the variance easily follows from Eq.~\ref{eq:sigma_omega_B}.

\subsection{Effect of Optical Intensity Noise}
Intensity noise in the optical waves in the proposed experiment transforms into amplitude noise in the BD's output current, from which we extract the Larmor frequency and hence the magnetic field strength.

Optical intensity noise is quantified as RIN. Frequency-resolved RIN is the power spectral density of the  intensity noise, normalized by the square of the intensity (i.e., the mathematical power of the intensity signal)~\cite{Keller2022}: 
\begin{equation} \label{eq:RIN_def}
    \mathcal{R}_f=\frac{\sigma_I^2}{\overline{I}^2}\frac{1}{ \Delta f},
\end{equation}
where $\overline{I}$ is the mean optical intensity, $\sigma_I^2$ is the intensity noise variance, and $\Delta f$ is the bandwidth, where we assume that noise is evenly distributed over frequency. 

The noise variance of the photocurrent resulting from detection of light with intensity noise variance $\sigma_I^2$ is given by~\cite{Ku1966}
\begin{equation}
        \sigma_{i}^2 = \left.\frac{di}{dI}\right|^2_{I=\overline{I}} \sigma_{I}^2= \overline{i}^2 \mathcal{R}_f \Delta f,
\end{equation}
where $\overline{i}$ is the mean photocurrent, we used the fact that $i \propto I$ and we assumed that $\sigma_i^2 \ll \overline{i}^2$ (i.e., reasonably low noise). The corresponding RMS noise spectral density is 
\begin{equation}
    \mathcal{N}_0^2 = \frac{\sigma_i^2}{\Delta f} = \bar{i}^2 \mathcal{R}_f.
\end{equation}
We then find
\begin{equation}
\sigma_{\omega_B}^2= \frac{6 \overline{i}^2 \mathcal{R}_f \left( e^{2T_m/T_2} - 1 \right) T_2}{e^2\epsilon_d^2\bar{n}_f^2 P_\text{pol}^2 T_m^4},
    \label{eq: sigma_AN}
\end{equation}
where we used Eq. \ref{eq:BD_output} to get $\left|\text{max}(i_{BD})\right| = e\epsilon_d\bar{n}_fP_\text{pol}$.

We note that, generally, we can write RIN as the sum of contributions from shot noise and technical noise, $\mathcal{R}_f = \mathcal{R}_{f, SN} + \mathcal{R}_{f, tech}$. Shot noise sets the minimum value of RIN, given by
\begin{equation} \label{eq:RIN_shot_noise}
    \mathcal{R}_{f,SN} = \frac{2 \hbar \omega}{P},
\end{equation}
where $\hbar$ is the reduced  Planck's constant and $P$ is the mean optical power (this result stems directly from the Poisson statistics of photon shot noise, where the variance in the number of photons, $n$, is $\sigma^2_n = n$, and correspondingly $\sigma^2_{I} = \left( dI/dn \right)^2 \sigma^2_n = I^2/n$; the factor of 2 compensates for the mathematical use of a one-sided spectrum). In some cases, one contribution dominates over the other, as we explain when we examine each noise source below.

\subsection{Fundamental Noise}
\label{sec:fund-noise}
Fundamental noise sources are unavoidable. They can only be mitigated by choice of operational parameters, but never eliminated through any technical method. We consider two fundamental noise sources, photon shot noise and spin projection noise.

\subsubsection{Photon Shot Noise}
\label{sec:shot-noise}
We separately analyze noise in the infrared (IR) light field and two UV laser fields used in this experiment. Recall that RIN generally contains both the fundamental contribution of shot noise and additional noise from technical sources. For the low-power fluoresced IR and UV reference fields that are incident onto the BD photodiodes, power is low enough for shot noise to be significant. For the high-power UV light that drives the two-photon excitation, intensity noise is dominated by technical noise. For example, consider UV laser power of $\SI{1}{W}$ (the lowest power we take into account). The shot noise contribution to RIN would be $\SI{-178}{dBc/Hz}$. The RIN values that we consider for this laser (as described below)  are between $\SI{-120}{dBc/Hz}$ and $\SI{-80}{dBc/Hz}$, i.e., at least $10^6$ times higher than the shot noise contribution. Therefore, we neglect the impact of shot noise of the high-power UV excitation light. On the other hand, the estimated detected UV power ranges from $\SI{16}{\micro W}$ to $\SI{1.5}{mW}$, for which the shot noise contribution to RIN would be $-130$ to $\SI{-149}{dBc/Hz}$.  Our experimental proposal relies on a balanced detector with common-mode rejection ratio of $\SI{45}{dB}$, which reduces the impact of technical RIN on the detected signal to a level comparable to the impact of shot noise (see Fig.~\ref{fig:w0_calculation}).  Therefore, shot noise in the detected UV light cannot be neglected.

By design, the detected UV reference power and corresponding RMS photocurrent are set to 
$\epsilon_q P_{r} = \epsilon_q e\epsilon_g \bar{n}_f / R(\omega_{UV}) = \epsilon_d\hbar\omega \bar{n}_f $ and $i_r = e\epsilon_d\bar{n}_f$, respectively, as explained in Section \ref{sec:basic_model}. From Eq. \ref{eq:RIN_shot_noise}, the shot noise contribution to RIN corresponding to this optical power is $\mathcal{R}_{f,r, SN} = 2 / \epsilon_d\bar{n}_f$. Putting these current and RIN values in Eq. \ref{eq: sigma_AN} we get the contribution of the reference UV shot noise to the error,
\begin{equation} \label{eq: UV_shot_noise}
    \sigma_{\omega_B, r, SN}^2 = \frac{12\left(e^{2T_{m}/T_{2}}-1\right)T_{2}}{\epsilon_{d}\bar{n}_{f}P_{\text{pol}}^{2}T_{m}^{4}}.
\end{equation}

We obtain the contribution of the IR shot noise to the error in the same manner, where $\overline{i}_{IR} = e\epsilon_d\bar{n}_f$. For the shot noise RIN we use the mean detected IR power $\epsilon_q \bar{P}_{IR} = \hbar \omega _{IR}\epsilon_d \bar{n}_{f}$, yielding $\mathcal{R}_{f, IR, SN} = 2/\epsilon_{d}\bar{n}_{f}$. Putting these into Eq. \ref{eq: sigma_AN} we get
\begin{equation} \label{eq: IR_shot_noise}
    \sigma_{\omega_B, IR, SN}^2 = \frac{12\left(e^{2T_{m}/T_{2}}-1\right)T_{2}}{\epsilon_{d}\bar{n}_{f}P_{\text{pol}}^{2}T_{m}^{4}}.
\end{equation}

Overall, the sensitivity due to shot noise alone is
\begin{equation}\
    \mathcal{S}_{B, SN} = \sqrt{\frac{T_m}{N_m}}\frac{\sqrt{\sigma^2_{\omega_B, r, SN} + \sigma^2_{\omega_B, IR, SN}}}{\gamma_{Xe}}.
\end{equation}

\subsubsection{Spin Projection Noise}
\label{sec:spin-projection}
The variance due to spin projection noise is \cite{Budker2013}
\begin{equation}
    \sigma_{\omega_B,PN}^2 = \frac{(2\pi)^2}{T_2 N_{\text{Xe}} T_m}.
    \label{eq: sigma_PN}
\end{equation}
It is independent of laser power and depends on the number of xenon atoms involved in the interaction, spin polarization decay time, and measurement time. The sensitivity due to spin projection noise alone is
\begin{equation}
    \mathcal{S}_{B, PN} = \sqrt{\frac{T_m}{N_m}}\frac{\sigma_{\omega_B, PN} }{\gamma_{Xe}}.
    \label{eq:sensitivity-PN}
\end{equation}

For most of the analysis in the remainder of the paper, we group projection noise and shot noise into a single category of fundamental noise. We then define the sensitivity due to fundamental noise as
\begin{equation}
    \mathcal{S}_{B, FN} = \sqrt{\mathcal{S}_{B, PN}^2 + \mathcal{S}^2_{B, SN}}.
\end{equation}

\subsection{Technical Noise}
\label{sec:tech-noise}

In the proposed experiment, there are two primary technical noise sources: laser intensity noise (technical RIN) and noise in the detection apparatus.

\subsubsection{Laser Intensity Noise}
\label{sec:rin}
The intensity noise of the high-power UV laser propagates to the detected Larmor frequency in two ways. First, through the low-power UV reference detected by the BD. Second, through the variation it induces in the two-photon excitation rate, which in turn induces variation in the detected IR photon rate.

To find the impact of the UV reference technical RIN, we simply use Eq. \ref{eq: sigma_AN} with $\overline{i}_{UV}= e\epsilon_d\bar{n}_f$ and set $\mathcal{R}_f = \mathcal{R}_{f,tech}$, which we keep as a free parameter, to get
\begin{equation} \label{eq:UV_tech_noise}
\sigma_{\omega_B, r, tech}^2 = \frac{6 \mathcal{R}_{f,tech} \left( e^{2T_m/T_2} - 1 \right) T_2}{P_\text{pol}^2 T_m^4 }.
\end{equation}

For the impact of the UV excitation light's technical noise on the detected IR, we need to take into account their nonlinear relation through two-photon excitation (the RIN in Eq. \ref{eq: sigma_AN} is the RIN in the detected light, not in the excitation light). From Eqs.~\ref{eq:n_f(t)} and \ref{eq:i_IR} we see that $I_{IR} \propto n_f $ and from Eqs.~\ref{eq:n_f} and \ref{eq:W2} we have $n_f\propto I_e^2$. Therefore,
\begin{equation}
    \sigma_{I_{IR}}^2 = \left( \frac{dI_{IR}}{dI_e} \right)^2 \sigma_{I_e}^2 = 4\frac{I_{IR}^2}{I_e^2} \sigma_{I_e}^2.
\end{equation}
Using this in Eq. \ref{eq:RIN_def} we get the resulting RIN in the IR light, $\mathcal{R}_{f} = \frac{4}{I_e^2 \Delta f} \sigma_{I_e}^2 = 4 \mathcal{R}_{f,tech}$, which we can now use in Eq. \ref{eq: sigma_AN} together with $\overline{i}_{IR} = e\epsilon_d\bar{n}_f$, yielding
\begin{equation}\label{eq:IR_tech_noise}
    \sigma_{\omega_B, IR, tech}^2 = \frac{24  \mathcal{R}_{f, tech} \left( e^{2T_m/T_2} - 1 \right) T_2}{P_\text{pol}^2 T_m^4 }.
\end{equation}

Here, we implement balanced detection to mitigate the effect of RIN on sensitivity. This mitigation is quantified through the BD's common mode rejection ratio (CMRR), $R_{CMRR}$, expressed as a multiplicative factor that reduces the impact of intensity noise as long as it is common to both of the BD's photodiodes. We include the CMRR in the variance due to RIN, defined as
\begin{equation}
    \mathcal{S}_{B,RIN}  =\sqrt{\frac{T_m}{N_m}}\frac{\sqrt{\sigma_{\omega_B, r, tech}^2 + \sigma_{\omega_B, IR, tech}^2}}{\gamma_{Xe} R_{CMRR}}.
    \label{eq:sigma_RIN}
\end{equation}

\subsubsection{Photodiode Detector Noise}
\label{sec:detector-noise}
The detector has frequency-dependent responsivity that has a maximum value of $R_{max}=R(\omega_{max})$. It is also characterized by the noise-equivalent power that has a minimum value of $NEP_{min}=NEP(\omega_{max})$. For any angular frequency $NEP(\omega)R(\omega) = NEP_{min}R_{max}$.
Therefore, the RMS noise spectral density for both the IR and low-power UV photocurrents is equal to $\left( NEP_{min}R_{max} \right)^2$, bringing the total RMS noise spectral density amplitude in the BD output current to $\mathcal{N}_{0}^2 =  2\left(NEP_{min} R_{max}\right)^2$. We use this value in the first expression in Eq. \ref{eq: sigma_AN} together with $\left|\text{max}(i_{BD})\right| = e\epsilon_d\bar{n}_fP_\text{pol}$ and get
\begin{equation}
    \sigma_{\omega_B,DN}^2 = \frac{12 \left(NEP_{min} R_{max}\right)^2 \left( e^{2T_m/T_2} - 1 \right) T_2}{e^2\epsilon_d^2\bar{n}_f^2 P_\text{pol}^2 T_m^4}.
   \label{eq:sigma_DN}
\end{equation}
We can then define the sensitivity due to detector noise as
\begin{equation}\label{eq:detector_noise}
    \mathcal{S}_{B,DN} = \sqrt{\frac{T_m}{N_m}}\frac{\sigma_{\omega_B, DN}}{\gamma_{Xe}}.
\end{equation}
\subsubsection{Detector Dark noise}
Thorlabs has performed a noise analysis on the BD under consideration here. The test was performed with the photodiodes covered to determine the noise due to the BD amplification independent of the incident light. Unfortunately, in the relevant frequency regime described below, the noise floor of the measurement system dominates. Therefore, we conservatively take the highest noise reported at $\sigma_{P,elec} = \SI{-40}{dBm} = \SI{1e-7}{W}$, which  was taken near DC with a bandwidth of $\Delta f = \SI{1}{kHz}$. We find that the RMS noise spectral density of a measurement taken across a $R_{elec}=\SI{1}{M\Omega}$ resistor as 
\begin{equation}
    \mathcal{N}_0^2 = \frac{\sigma_{i_{BD}}^2}{\Delta f} = \frac{1}{R_{elec}^2\overline{i}_{BD}^2}\frac{\sigma_{P, elec}^2}{\Delta f},
\end{equation}
where the RMS of the BD output current is $i_{RMS} = e\epsilon_d\bar{n}_f P_\text{pol}/\sqrt{2}$ and we used $\sigma_{i}^{2}=\left({di}/{dP_{elec}}\right)^{2}\sigma_{P_{elec}}^{2} = {1}/{i^{2}R_{elec}^{2}}\sigma_{P_{elec}}^{2}$. The resulting contribution to $\sigma_{\omega_B}^2$ is six orders of magnitude lower than any other contribution and is neglected.

\section{Parameter Values} \label{sec:parameters_opt}

\subsection{Parameters values from Literature}\label{sec:parameters_lit}

\begin{table*}[htbp]
\centering
    \caption{Summary of Values}
    \label{tab:values}
    \begin{tabular}{cccc}
         Parameter& Symbol & Value & Reference\\
         \hline Initial polarization&$P_{\text{pol}}$&0.90\,\si{}&\cite{Nikolaou2013}\\
Isotope concentration&$\eta_{{^{{129}}\text{Xe}}}$&0.80\,\si{}&-\\
Xenon pressure&$p_\text{Xe}$&1\,\si{mbar}&\cite{Kilian2007, Nikolaou2013}\\
Xenon temperature&$T_\text{Xe}$&293\,\si{K}&\cite{Kilian2007}\\
Pulse length&$\tau_c$&66.7\,\si{ps}&\cite{ozawa2012}\\
Two-photon absorption coefficient&$\alpha_\text{Xe}$&83\,\si{cm^4/J^2}&\cite{kroll1990}\\
Natural linewidth&$\gamma_{nat}$&4.3\,\si{MHz}&\cite{Whitehead1995}\\
Pressure broadening&$\gamma_{pb}$&18.6\,\si{MHz/mbar}&\cite{Raymond1984}\\
Gyromagnetic ratio&$\gamma_\text{Xe}/2\pi$&11.78\,\si{MHz/T}&\cite{Fan2016}\\
Spin dephasing time&$T_2$&8000\,\si{s}&\cite{Kilian2007}\\
Photon collection efficiency&$\epsilon_g$&0.25\,\si{}&\cite{Cerez1990}\\
Max. detector responsivity&$R_{max}$&0.5\,\si{A/W}&Thorlabs PDB220A2\\
Detector noise equivalent power&$NEP_{min}$&3.6\,\si{pW/Hz^{1/2}}&Thorlabs PDB220A2\\
Common mode rejection ratio&$\mathcal{R}_{CMRR}$&45\,\si{dB}&Thorlabs PDB220A2\\
Measurement time&$T_m$&$\SIrange{1}{200}{min}$&-\\
Relative intensity noise&$\mathcal{R}_f$&$\SIrange{-80}{-120}{dBc/Hz}$&\cite{Gao2023,Luo2020,Xiu2023,Shestaev2020,Li2016}\\
Laser power&$P_l$&$\SIrange{1}{100}{W}$&\cite{Yang2012,Mueller2017}\\
Beam waist radius&$w_0$&$\SIrange{0.1}{10.0}{mm}$&-\\
     \end{tabular}
\end{table*}

The expression for sensitivity depends on many operational parameters. We begin by setting some of them to conservatively estimated values that are experimentally feasible. We begin by noting that $T_2$ times as long as 8000 seconds have been demonstrated in a 4:1 xenon-nitrogen gas mix with xenon partial pressure of about $\SI{50}{mbar}$, where lower pressure correlated with longer $T_2$ \cite{Kilian2007}. Here we conservatively assume $T_2 = \SI{8000}{s}$ for a pressure of $\SI{1}{mbar}$. The xenon gas is assumed to be at room temperature, $\SI{293}{K}$. We conservatively assume $\eta_{^{129}\text{Xe}} = 0.8$, though isotopically enriched xenon gas with significantly more than 80\% $^{129}\text{Xe}$ content is commercially available. Methods for spin polarization of $^{129}\text{Xe}$ have been thoroughly explored and established \cite{Walker1997}. For example, spin-exchange optical pumping has been shown to deliver a polarization fraction of 90\% for a gas pressure of $\SI{400}{ mbars}$ \cite{Nikolaou2013} with lower pressures associated with higher polarization fractions. We conservatively assume $P_\text{pol} = 0.9$ for xenon pressure of 1 mbar, though a higher value is likely possible at the pressure considered here. We emphasize that no alkali vapour is present in the detection phase, even if it has previously been used for hyperpolarization, as we explained in Section~\ref{sec:operation}. 

The infrared driving laser is assumed to be a Yb:fiber frequency comb laser system capable of over 100 Watt average power.  This laser is frequency-quadrupled via two second-harmonic generation (SHG) processes to provide the 256 nm UV light for two-photon excitation in xenon, where an overall conversion efficiency of a few percent is expected \cite{Yang2012, Mueller2017}. See Appendix~\ref{app:laser} for further details on laser selection. While short infrared pulses are vital for high conversion efficiencies, using longer (chirped) UV excitation pulses does not alter the two-photon excitation rate while allowing for a larger interaction volumes~\cite{ozawa2012}. Therefore, we take the UV pulse length to be $\tau_{pulse} = \SI{67}{ps}$, which corresponds to an interaction length of $L = c\tau_{pulse}=\SI{2}{cm}$.

The sensitivity depends on the efficiency of the detection of IR photons. As outlined above, this is divided into collection and quantum efficiency. The collection efficiency of the emitted near-infrared photons depends on the geometry of the experiment, and standard methods commonly achieve $\sim 50\%$ collection efficiency~\cite{Cerez1990}; in the below calculations, we conservatively take $\epsilon_g=0.25$. The quantum efficiency is determined by the detector's specifications.

The BD is modeled after the Thorlabs PDB220A2. This model has a minimum noise equivalent power (NEP) of $NEP_{min} = \SI{3.6}{pW/\sqrt{Hz}}$, maximum responsivity of $R_{max} = \SI{0.5}{A/W}$,  and a CMRR of $\mathcal{R}_{CMRR} \approx\SI{45}{dB}$  for frequencies below $\SI{1}{MHz}$.  Using equation~\ref{eq:responsivity} we get the quantum efficiency for the IR and UV, $\epsilon_q(\omega_{IR}) = 0.68$ and $\epsilon_q(\omega_{UV}) = 0.45$, respectively.

 Finally, we consider two properties of the xenon atom: the two-photon transition linewidth and the two-photon absorption coefficient. The natural linewidth of the two-photon transition is $\gamma_{nat} = \SI{4.3}{MHz}$ \cite{Whitehead1995}, and it is collisionally broadened at $\gamma_{pb} = \SI{18.6}{MHz/mbar}$ \cite{Raymond1984}, so that $\gamma = \gamma_{nat} + \gamma_{pb} \cdot p_\text{Xe}$. Since we are utilizing a DF configuration, Doppler broadening is not considered. The assumed $\SI{1}{mbar}$ pressure results in a linewidth of $\gamma=\SI{27.2}{MHz}$. Finally, the most recently reported value for the two-photon absorption coefficient is $\alpha_\text{Xe} = \SI{83}{cm^4/J^2}$~\cite{kroll1990}.

All values described above are summarized in Table~\ref{tab:values} together with the ranges of independent variables considered in Section \ref{sec:results}.

\subsection{Optimizing sensitivity through beam size}
\begin{figure}[htbp]
    \centering
    \includegraphics[width=\linewidth]{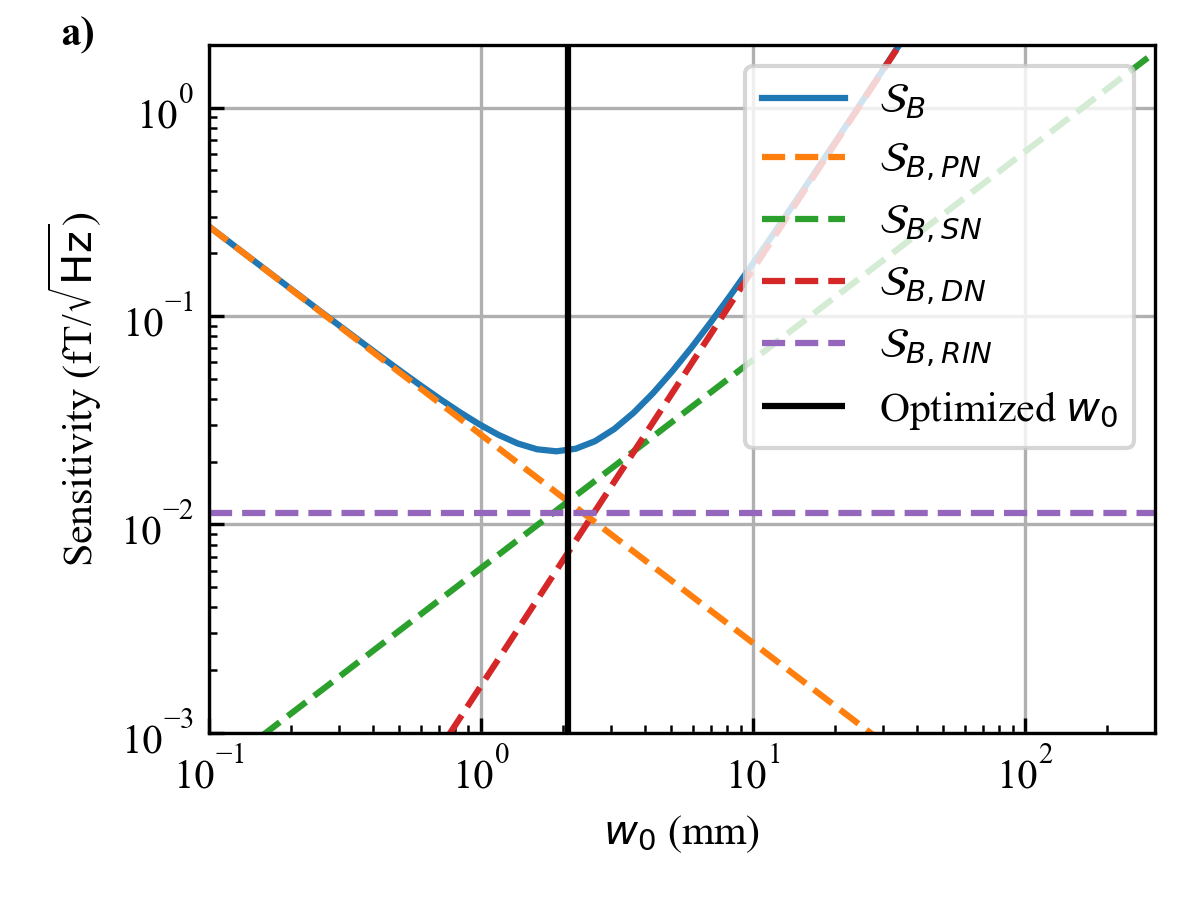}

    \caption{An example of beam size optimization, where $\mathcal{R}_f = \SI{-100}{dBc/Hz}$, $P_e = \SI{10}{W}$, and $T_m = 10$ min.  The dashed lines show dependence on beam size of the sensitivity due to projection noise (orange), shot noise (green), detector noise (red), and RIN (purple).  The blue solid line shows how the total sensitivity changes as a function of the beam waist, exhibiting a clear minimum. The corresponding optimal beam size is indicated with the black solid line.}
    \label{fig:w0_calculation}
\end{figure}

There are four parameters not addressed in the above section. Measurement time, laser RIN, and laser power are considered as independent variables in Section~\ref{sec:results}. The final parameter to discuss is the beam size, $w_0$. The dependence of contributions to the sensitivity on the beam size is shown in Figure~\ref{fig:w0_calculation}. Each source of noise in the experiment has a different response to changes in beam waist. Since RIN does not depend on beam size, neither does its contribution to the sensitivity, as shown by Equation~\ref{eq:sigma_RIN}. Projection noise is unique in that $\mathcal{S}_{B, PN}$ decreases as the beam waist increases, as shown in Equations~\ref{eq: sigma_PN} and ~\ref{eq:sensitivity-PN}. This happens because increasing the beam causes more xenon atoms to be illuminated and included in the interaction. The final two noise sources, shot noise and detector noise, increase as beam size increases. For a fixed laser power the average fluorescence scales as $\overline{n}_f \propto I^2 N_{Xe} \propto w_0^{-2}$. Thus, shot noise scales as $\mathcal{S}_{B,SN} \propto1/\sqrt{\overline{n}_f} \propto w_0$ and detector noise scales as $\mathcal{S}_{B, DN}\propto 1/\overline{n}_f \propto w_0^2$. 

The optimal beam size, that achieves the best sensitivity, is found by minimizing the quadrature sum of all noise sources. The details of this procedure are presented in Appendix~\ref{app:beam-waist} and an example is shown in Figure~\ref{fig:w0_calculation}. In the remaining calculations, the beam waist is optimized for the minimum noise unless the optimized value of the beam waist is outside the range $\SI{0.1}{m m}$-$\SI{10}{mm}$, which is practically easily achievable. If the optimized beam size was smaller or larger than these values, it was set to the closest allowed value.

\section{Results}
\label{sec:results}

We are left with three independent variables: measurement time, laser RIN, and laser power. We optimize the beam size for each case. Note that the fundamental noise and detector noise are independent of the laser technical RIN, and the laser technical RIN is independent of the laser power. 

We are interested in measuring sub-femtotesla magnetic fields. The Larmor frequency for such low fields is extremely low, $\ll \SI{1}{Hz}$, where high-power Yb:fiber laser RIN is in the range of -50 to -80 dBc/Hz but drops to -100 dBc/Hz or better at frequencies above 10 Hz \cite{Gao2023, Luo2020, Xiu2023, Shestaev2020, Li2016}. Therefore, the impact of RIN can be mitigated by shifting the measured signal to a higher frequency. This is often done in highly sensitive alkali magnetometers as well \cite{Shah2007} to avoid 1/f noise, e.g., by applying a weak ($\sim \SI{}{pT}$) oscillating magnetic field. Methods of laser RIN reduction,  suggested or applied to different types of lasers, could also significantly reduce RIN in this case. We refer the reader to Li et al. \cite{Li2024} and Ebrahimzadeh et al \cite{Ebrahimzadeh2025} for a concise summary of such methods and a proposal for a novel one based on SHG. Therefore, we consider RIN in the range -80 to -120 dBc/Hz to be feasible here.

\begin{figure*}[htbp]
    \centering
    \includegraphics[width=0.3\linewidth]{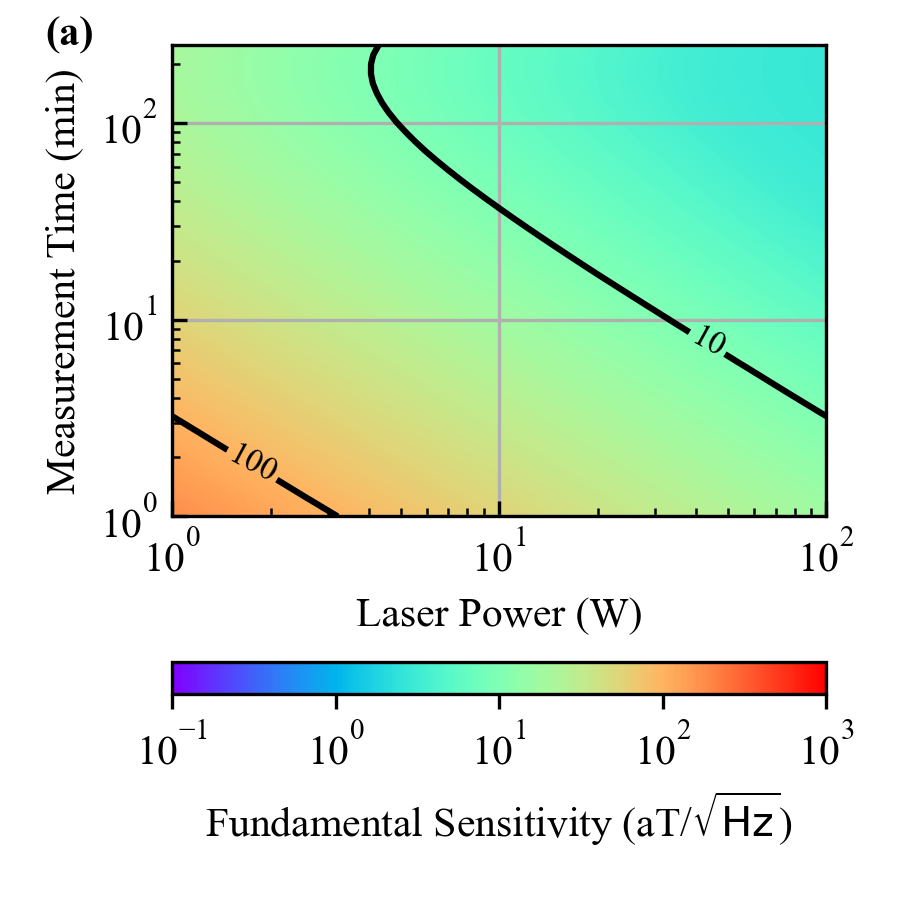}
    \includegraphics[width=0.3\linewidth]{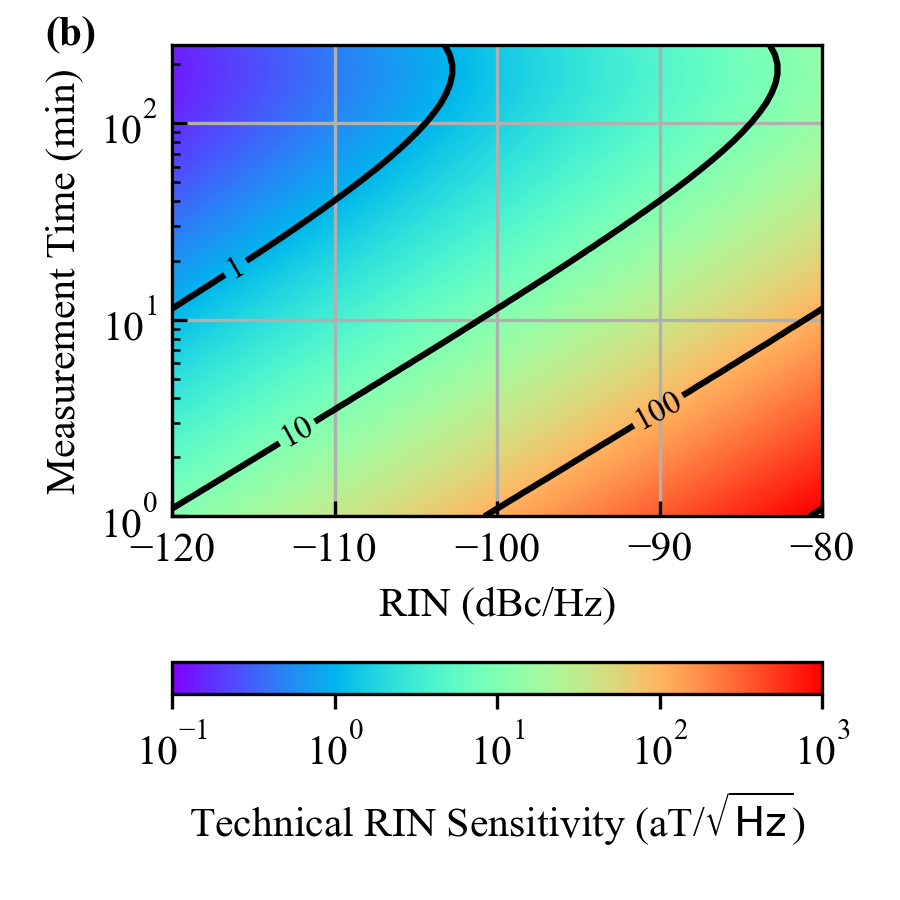}
    \includegraphics[width=0.3\linewidth]{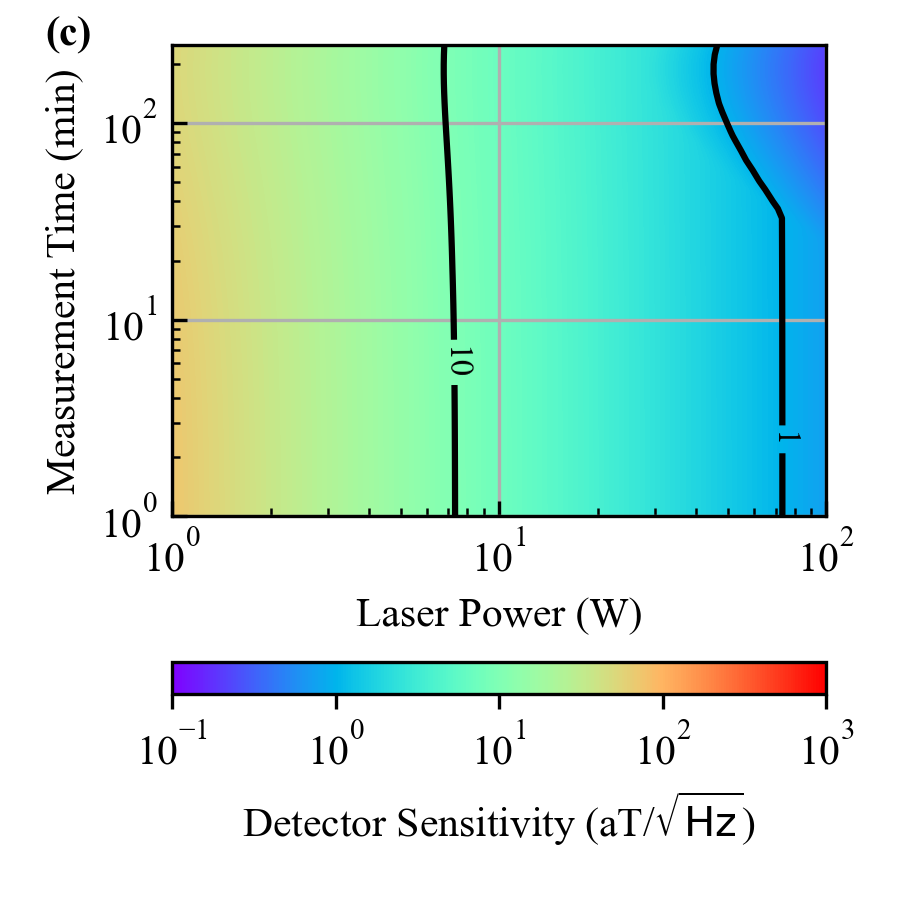}
    \caption{Contour plots showing each type of noise contribution to the sensitivity. In all plots the color map has the same range with the order of magnitude levels marked in black for increased clarity. a) Sensitivity due to fundamental noise, i.e., shot and projection noise. b) Sensitivity due to laser technical RIN. c) Sensitivity due to detector noise. }
    \label{fig:contour-sensitivity}
\end{figure*}

In Fig.~\ref{fig:contour-sensitivity}, each contour plot shows how each contribution depends on independent parameters values (for $N_m=1$). In all three plots, as the measurement time increases, the sensitivity improves until $T_m \sim \SI{200}{min}$, where  the signal has decayed so that further measurement does not improve sensitivity. Figure~\ref{fig:contour-sensitivity} (a) shows the sensitivity due to the fundamental noise sources. As the laser power increases, the fundamental sensitivity improves since shot noise is decreasing. The contribution of laser technical RIN is shown in Figure~\ref{fig:contour-sensitivity} (b). As RIN decreases, the sensitivity improves. Finally, the contribution of detector noise to sensitivity is shown in Figure~\ref{fig:contour-sensitivity} (c). It is significantly lower than the fundamental noise contribution for all values of laser power and measurement time considered here.
\begin{figure}[htbp]
    \centering
    \includegraphics[width=1\linewidth]{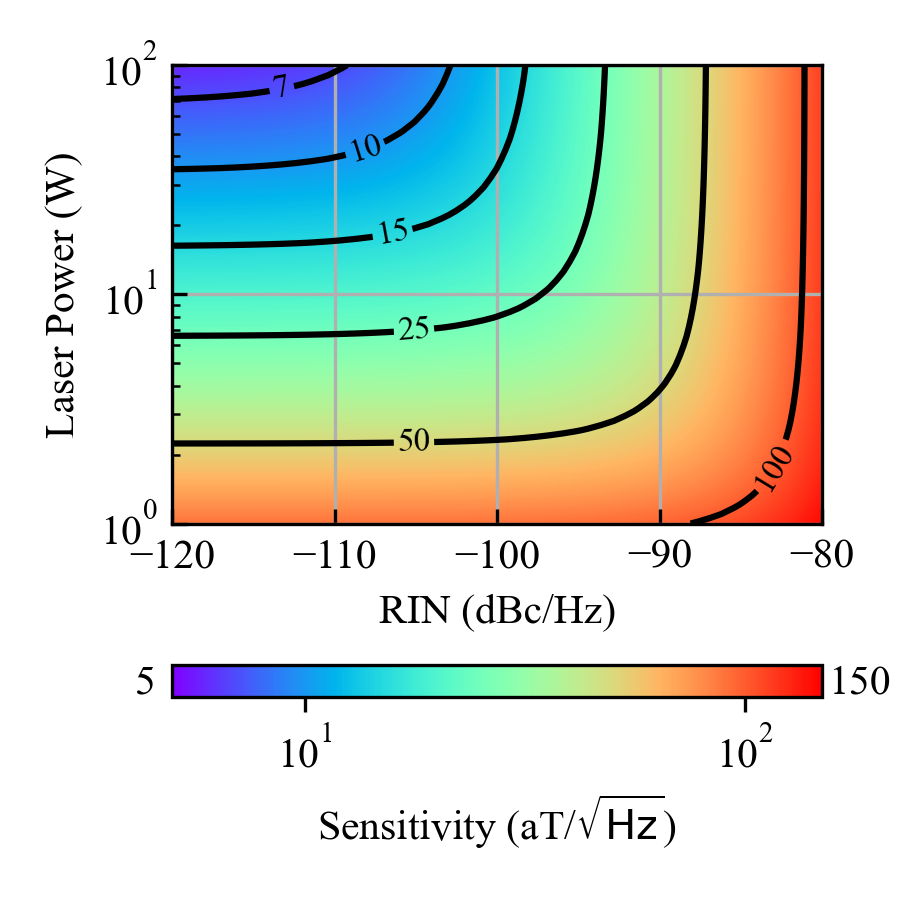}
    \caption{Contour plot showing the effect of laser power and technical RIN on sensitivity with a measurement time of 10 minutes.  Note that the color map in this plot has a different scale than those in Figure~\ref{fig:contour-sensitivity}.} 
    \label{fig:final-plot}
\end{figure}
\begin{figure}[htbp]
    \centering
    \includegraphics[width=1\linewidth]{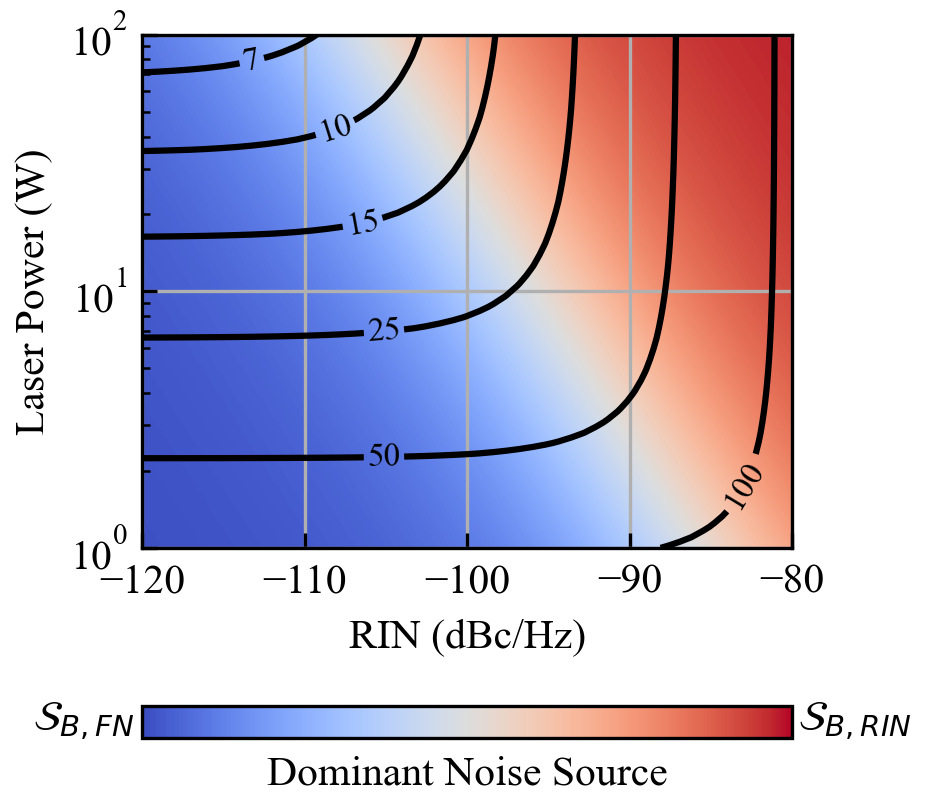}
    \caption{The color plot shows the dominant noise source  vs. laser power and technical RIN, as defined by $\mathcal{D}$ in Eq. \ref{eq:dominant}. $\mathcal{D} < 0$ values show that fundamental noise dominates (in blue), and $\mathcal{D}>0$ values show that RIN dominates (in red). The black contour lines are the same as those in Figure~\ref{fig:final-plot}, for reference.}
    \label{fig:final-plot-ratio}
\end{figure}

With an understanding of how each noise contribution varies over measurement time and with laser power or RIN, we aim to analyze possible measurement strategies. Figure~\ref{fig:contour-sensitivity} shows that, in general, increasing the measurement time improves sensitivity (up to $\SI{200}{min}$). In Figures~\ref{fig:final-plot} and~\ref {fig:final-plot-ratio} we analyze the sensitivity and its limiting factor for a single (i.e., $N_m=1$) measurement time of 10 minutes, which is easily feasible. Figure~\ref{fig:final-plot} shows the total sensitivity for a range of laser power and RIN values. The colormap in this plot is different from that in Figure~\ref{fig:contour-sensitivity}. The worse sensitivity is present for larger RIN and lower laser power, as we expect.  The shape of the contour lines in Figure~\ref{fig:final-plot} indicates that increasing the power of the UV excitation laser does not always substantially improve the sensitivity. This happens when RIN is the dominant source of noise. 
  To investigate this we define the function 
  \begin{equation} \label{eq:dominant}
  \mathcal{D}(P_e, \mathcal{R}_{f, tech}) = \frac{\mathcal{S}_{B, RIN}-\mathcal{S}_{B, FN}}{\sqrt{\mathcal{S}_{B,RIN}^2 + \mathcal{S}_{B,FN}^2}},
  \end{equation}
  where $\mathcal{D} < 0$ values show that fundamental noise dominates, and $\mathcal{D}>0$ values show that technical noise dominates (recall that the technical sensitivity does not depend on the laser power and the fundamental noise does not depend on the technical RIN). For the ranges considered here the detector noise is much lower than both RIN and fundamental noise, and is therefore neglected. In Figure~\ref {fig:final-plot-ratio} the ranges of laser power and RIN, and the black contour lines, are the same as in Figure~\ref{fig:final-plot}. The colour plot shows $\mathcal{D}$ and thus indicates which noise type is dominant: red indicates that the technical noise sources are dominant, and blue means that the fundamental noise sources dominate. 
  
  There are three clear regions in which different approaches for improving the sensitivity are most effective. The first region is in the top right side of Figure~\ref{fig:final-plot-ratio}, which is dominated by RIN, and the optimal method for improving the sensitivity is to reduce RIN. The second region is in the bottom-left corner, which is dominated by fundamental noise, and the optimal method to improve sensitivity is to increase the laser power. And finally, the area between the two clear regions demarks a region where improving either the laser power or RIN will result in improving the sensitivity. Therefore, this plot provides important guidance for improving the sensitivity of a given system.

We highlight three exemplary situations: realistic with current technology, optimal with current technology, and  optimal with near-future technology. For the case that is realistic with current technology, we take a $\SI{1}{W}$ laser with a RIN of $\SI{-80}{dBc/Hz}$, resulting in a  sensitivity of $\SI{148}{aT/\sqrt{Hz}}$, which is slightly better than the current limit set by SQUIDS of $\SI{150}{aT/\sqrt{Hz}}$. This case is dominated by technical noise, where the optimal method for improving the sensitivity is to reduce RIN. Next, a $\SI{10}{W}$ laser (possible in the near future) and $\SI{-100}{dBc/Hz}$ RIN (currently possible with modulation)  would result in a sensitivity of $\SI{24}{aT/\sqrt{Hz}}$, which is more than 6 times better than the current sensitivity record. The noise in this situation is approximately evenly distributed between technical and fundamental noise sources. Improving either laser power or RIN would improve the sensitivity. Next, we consider the sensitivity that can be expected in the near future. A $\SI{100}{W}$ laser with a RIN of $\SI{-120}{dBc/Hz}$ could produce a sensitivity of $\SI{5}{aT/\sqrt{Hz}}$, which is 30 times better than the current record. This limit is bounded by fundamental noise, where further improvements to the sensitivity are more dependent on laser power.

 By optimizing the beam waist and carefully considering the effects of both laser power and RIN, this  technique could exceed the current atomic and SQUID magnetometers record sensitivity using near-future technology. The primary technical consideration is the high-power, low-RIN UV laser.

\section{Conclusion}
We presented and analyzed a novel approach to quantum magnetic sensing using optical detection of noble gas nuclear spin precession via two-photon excitation, where the light source is a frequency-quadrupled high-power fibre frequency comb laser system, and using balanced detection. Our analysis shows that, under realistic experimental conditions, fundamental and relative intensity noise will dominate the  magnetometer's sensitivity. We found that optimizing the beam waist is vital for optimizing the sensitivity due to fundamental noise, and that using a BD significantly improved the sensitivity due to technical noise.  Depending on which noise source is dominant, the technique for further improving sensitivity differs: increased laser power is required when the fundamental noise dominates, and improved RIN when technical noise dominates. We found that exceeding state-of-the-art performance would require an ultraviolet laser of $\SI{1}{W}$ combined with RIN of $\SI{-80}{dBc/Hz}$ and a measurement time of 10 min. With a $\SI{10}{W}$ laser and a RIN of $\SI{-100}{dBc/Hz}$ we predict magnetometric sensitivity of $\SI{24}{aT/\sqrt{Hz}}$ is feasible. 
\begin{acknowledgments}
We are grateful to Prof. Lilian Childress for advice on balanced detection. We acknowledge the support of the Natural Sciences
and Engineering Research Council of Canada (NSERC),
[funding reference numbers RGPIN-2019-05017], NSERC Alliance Consortia Quantum Grant AQUA (ALLRP
587602-23), NSERC Alliance-Alberta Innovates Advance [570922-21 and 212200789], National Research Council (NRC) Canada Quantum Sensors Challenge Program [CSTIP Grant \#QSP-108-1], University of Alberta Faculty of Engineering Research Exploration Fund.
\end{acknowledgments}
The data that support the findings of this article are openly available \cite{Maldaner2026}.
\appendix

\section{Two-photon selection rules}\label{app:two_photon_rules}
Here we provide details on selection rules for the two-photon transition~\cite{Bonin1984} considered in the main text. The fine structure ground state is $|g\rangle=5p^6(^1S_0)$ , and the magnetic sub-states present in $|g\rangle$ with an $F = 1/2$ are \mbox{$|\downarrow\rangle$} for $m_F=-1/2$ and $|\uparrow \rangle$ for $m_F = 1/2$. The excited fine structure state is $5p^5(^2P_{3/2})6p^2[5/2]_2$ and the magnetic sub-state with $F^\prime=3/2$ is $|e\rangle$ with $m_{F^\prime}=3/2$.

The requirements for an allowed two-photon transition are summarized in Table~\ref{tab:two-photon}, along with a verification for the transitions considered here. Due to the requirement for $\Delta m_F = 2$ only the $|\downarrow \rangle$ state has an allowed transition to $|e\rangle$ while $|\uparrow\rangle$ to $|e\rangle$ is forbidden.

\begin{table}[htbp]
    \centering
    \caption{Two-photon transition requirements}
    \label{tab:two-photon}
    \begin{tabular}{l|l|l}
        Rule & Our Values & Satisfied? \\
        \hline
        \hline
         $|\Delta F| \leq 2$& $F - F^\prime=1$  & Yes \\
         \hline
         Same Parity& $|g\rangle$: Even, $|e\rangle$: Even & Yes \\
         \hline
         $F + F^\prime \in \mathbb{N}$& $F + F^\prime = 2$ & Yes \\
         \hline
         \multirow{4}{11em}{$\Delta F$ not one of: \\$\qquad0\leftrightarrow0$\\ $\qquad1\rightarrow1$\\ $\qquad \sfrac{1}{2}\rightarrow \sfrac{1}{2}$}&&\\
         &&\\
         &&\\
         &$ \sfrac{1}{2}\rightarrow \sfrac{3}{2}$&Yes\\
         \hline
         \multirow{2}{11em}{$\Delta m_F=2$}&$|\downarrow\rangle$: $m_F - m_{F^\prime} = 2$&Yes\\
         &$|\uparrow\rangle$: $m_F - m_{F^\prime} = 1$&No\\
    \end{tabular}
\end{table}

\section{Laser Choice}
\label{app:laser}
Our choice of a frequency comb laser warrants some discussion. The UV laser field is assumed to consist of a single TEM$_{00}$ transverse mode, be wavelength-tuned to $\SI{256}{nm}$ with MHz precision, have linewidth well below $\SI{20}{MHz}$, and $\SIrange{1}{100}{W}$ of average power. Such direct lasing at $\SI{256}{nm}$ is extremely challenging, mainly due to the absence of appropriate gain media. Therefore, we propose frequency quadrupling an infrared driving laser at $\SI{1024}{nm}$, where excellent laser technology is readily available. We consider two types of infrared driving lasers that could satisfy our scheme's spectroscopic precision requirements: CW  and frequency comb lasers.

CW lasers appear to be attractive due to their narrow linewidth and simple spectrum. However, a main consideration is the efficiency of the second SHG step, which is commonly rather low due to constraints arising from the properties of the few technologically mature nonlinear crystals that are transparent in the deep UV, particularly low nonlinear susceptibility. SHG is a nonlinear process, thus higher intensity yields higher efficiency, and a lower nonlinear susceptibility further increases the necessary intensity. This poses a significant challenge for CW lasers, resulting in commonly low SHG efficiency in the deep UV. Moreover, amplification of single-transverse-mode CW lasers to high power requires GHz-scale broadening of the linewidth to avoid detrimental nonlinear effects~\cite{Liu2024}. To the best of our knowledge, only recently was Watt-level deep UV CW power achieved in this manner~\cite{TerMikirtychev2026}.  A CW fiber laser, at $\SI{1030}{nm}$ with $\SI{376}{W}$ of power and a linewidth of $\SI{9.6}{GHz}$, generated $\SI{0.91}{W}$ of $\SI{257}{nm}$ light . A similar laser, with $\SI{1.5}{kW}$ power at $\SI{1064}{nm}$ and a linewidth of $\SI{60}{GHz}$, generated $\SI{16}{W}$ at $\SI{266}{nm}$. The efficiency of the second SHG process was $\sim\SI{1}{\%}$ and $\sim \SI{2}{\%}$, respectively, and anyway the linewidths are far too large for our scheme. These issues are partially circumvented by using intra-cavity configurations, where the driving laser power can build up to a very high value. Experimental demonstrations reach $\SIrange{1.4}{3}{W}$ at $\SIrange{243}{266}{nm}$ with linewidth as low as $\SI{10}{kHz}$, where the infrared driving laser power was $\SIrange{7}{42}{W}$, and the efficiency of the second SHG process was $\SIrange{30}{40}{\%}$~\cite{Burkley2019, Li2026, Sun2026, Burkley2021}. Notably, Sudemeyer et. al.~\cite{Sudmeyer2008}  were able to achieve $\SI{12}{W}$ at $\SI{266}{nm}$ (with $\SI{57}{\%}$ conversion efficiency for the second SHG process), but only for a limited time due to degradation of cavity optics under high-power visible and deep UV illumination, which caused the deep UV power to drop to $\SI{4.5}{W}$. While these achievements are impressive, and potentially could benefit the magnetometric scheme proposed here, this is only true up to a point, as further power scaling (towards $\SI{100}{W}$) seems daunting not only for the narrowband infrared pump, but also due to cavity stability and optical damage limitations.

Contrary to CW lasers, a frequency comb laser produces a stabilized train of pulses of femtosecond-to-picosecond duration, at MHz-GHz repetition rate. Therefore, for the same average power, it provides much higher peak power, serving to greatly increase the efficiency of SHG. The spectrum of a frequency comb is composed of many narrow spectral features (``comb teeth"), each commonly having sub-MHz linewidth, which can be tuned with very high precision (typically kHz or better). For the special case of two-photon excitation, the entire spectrum of the frequency comb can be used: comb teeth that are equally distributed about half of the two-photon resonance frequency coherently add up pairwise to drive the same transition, effectively producing the same excitation as a CW laser with the same average power as the entire comb and the linewidth of a single comb tooth~\cite{Picque2019}.

Importantly, the excellent phase stability (and corresponding narrow tooth linewidth) of comb lasers is maintained even in amplification to very high power, thanks to the use of chirped-pulse amplification techniques that mitigate detrimental nonlinear effects~\cite{Schibli2008}. Frequency comb power has gradually been scaled up to $\SI{230}{W}$ with kHz linewidth~\cite{Ruehl2010, Li2016, Luo2020, Schmid2024}, using Yb:fiber  amplifiers operating at or near $\SI{1030}{nm}$. Similar Yb:fiber femtosecond laser systems have reached $\SI{1.2}{kW}$ with $\SI{1.39}{GHz}$ repetition rate with a single amplifier chain~\cite{Xiu2023}, and up to $\SI{10.4}{kW}$ of power using coherent combining of multiple amplifiers~\cite{Muller2018, Muller2020}, which relies on excellent phase stability, thus providing a strong indication that comb operation is feasible. This is further strengthened by the demonstration of 1 kW of power in a system with stabilized carrier-envelope offset~\cite{Shestaev2020}.  At present we are aware of only one demonstration of a fully stabilized deep UV frequency comb with Watt-level power~\cite{Yang2012}.  At $\SI{258}{nm}$, this comb’s power was $\SI{1.62}{W}$, and the second SHG process had $\SI{12}{\%}$ efficiency. Other notable works at conventional comb repetition rates ($>\SI{10}{MHz}$) reached $\SI{2.9}{W}$ at $\SI{266}{nm}$~\cite{Samanta2015} (though only $\SI{1}{W}$ with long-term stability) and $\SI{1.42}{W}$ at $\SI{256.7}{nm}$~\cite{Wang2025}, with $\SI{35}{\%}$ and $\SI{17}{\%}$ efficiency for the second SHG process, respectively. The infrared power available for these experiments was $\SI{100}{W}$, $\SI{20}{W}$, and $\SI{59}{W}$, indicating infrared-to-deep UV conversion efficiencies of $\SI{1.62}{\%}$, $\SI{14.5}{\%}$, and $\SI{2.4}{\%}$ respectively. The above advances in kW-power, phase-stable, high-repetition-rate Yb:fiber ultrafast laser systems, and the demonstrated generation of deep UV with few-percent efficiency, provide strong evidence that kW infrared frequency combs are feasible with existing technology, as is their use for generating deep UV combs with tens of Watts of power.

\section{Beam waist optimization}
\label{app:beam-waist}

We write the total variance as follows
\begin{equation}
    \sigma_{\omega_B}^2 = \sigma_{\omega_B,PN}^2 +\sigma_{\omega_B,SN}^2 +\sigma_{\omega_B,IN}^2 + \sigma_{\omega_B,DN}^2.
\end{equation}
We define the coefficients
\begin{align}
    A_{PN}&= \frac{(2\pi)^2}{T_2 T_m L\pi n_\text{Xe}} \\
    A_{SN}&=\frac{24 T_2 (e^{2T_m/T_2}-1)}{\sqrt{2}P_\text{pol}T_m^4 \frac{1}{2} \epsilon_d (L \pi n_\text{Xe})6 \alpha_{\text{Xe}} g(0)\frac{P_e^2}{\pi^2}}w_0^2\\
    A_{IN}&= \sigma_{\omega_B,IN}^2 \\
    A_{DN}&= \frac{6 (NEP_{min}R_{max} )^2(\exp(2T_m/T_2)-1)T_2}{(P_\text{pol}e)^2T_m^4(\frac{1}{2}L \pi n_\text{Xe}\epsilon_d 6\frac{P^2}{\pi^2}\alpha_{\text{Xe}} g(0))^2}w_0^4
\end{align}
and use Equations~\ref{eq: sigma_PN}, \ref{eq: IR_shot_noise}, \ref{eq: UV_shot_noise}, \ref{eq:IR_tech_noise} and \ref{eq:sigma_DN}, to write the variance in the Larmor frequency as
\begin{equation}
    \sigma_{\omega_B}^2= A_{PN} \frac{1}{w_0^2} + A_{SN}w_0^2 + A_{IN} + A_{DN} w_0^4,
\end{equation}
where we note that the variance due to RIN is independent of the beam size

We then find the minimum of $\sigma_{\omega_B}^2$ with respect to $w_0$ by finding where its derivative vanishes, i.e.,
\begin{align}
    \frac{d}{dw_0} \sigma_{\omega_B}^2 &= -2A_{PN}\frac{1}{w_0^3} + 2 A_{SN}w_0 + 4 A_{DN}w_0^3=0.
\end{align}
This can be written as
\begin{equation}
    0 = -2 A_{PN} + 2 A_{SN} w_0^4 + 4 A_{DN} w_0^6
\end{equation}
As this is a polynomial of degree 6, we solve it numerically using Python's SciPy fsolve function. 

To verify that this is a minimum rather than a maximum, we find the second derivative and ensure that it is positive as follows,
\begin{align}
    \frac{d^2}{dw_0^2}\sigma_{\omega_B}^2 &= 6 A_{PN}\frac{1}{w_0^4} + 2 A_{SN} + 12 A_{DN}w_0^2.
\end{align}
Since $A_{PN}$, $A_{SN}$, and $A_{DN}$ are positive, $\frac{d^2}{dw_0^2}\sigma_{\omega_B}^2 > 0$ for all values of $w_0$ and therefore the  $w_0$ value found above must be at a minimum of $\sigma_{\omega_B}^2$.

\end{document}